# Diffusion-Assisted Frequency Attention Model for Whole-body Low-field MRI Reconstruction

Xin Xie[1, #], Yu Guan[1, #], Zhuoxu Cui[2], Dong Liang[2], Qiegen Liu[3, *]

[1] *School of Mathematics and Computer Sciences, Nanchang University, Nanchang 330031, China. ({409100230013, 359100220011}@email.ncu.edu.cn)*

[2] *Research Center for Medical AI, Shenzhen Institute of Advanced Technology, Chinese Academy of Sciences, Shenzhen 518055, China. ({zx.cui, dong.liang}@siat.ac.cn)*

[3] *School of Information Engineering, Nanchang University, Nanchang 330031, China. (liuqiegen@ncu.edu.cn)*

[#] *These authors contributed to this work equally.*

*Corresponding author.*

**\*Correspondence to:**

Qiegen Liu, Ph.D.

School of Information Engineering, Nanchang University, Nanchang, China, 330031

Email: liuqiegen@ncu.edu.cn

Total word count: 5000 words

Number of Figures: 7

# Abstract

**Purpose:** A deep learning model is proposed to improve the reconstruction efficiency while enhancing the accuracy in the whole-body low-field magnetic resonance imaging (MRI) reconstruction.

**Methods:** Although diffusion model (DM) has demonstrated great potential in conventional MRI, the direct application to low-field MRI remains challenging due to the inherently weak signal and low signal-to-noise ratio (SNR) associated with low-field imaging. These limitations often lead to performance degradation and exacerbate the reconstruction inaccuracies observed in traditional methods. To address these issues, we propose a diffusion-assisted frequency attention model (DFAM), which integrates the generative capabilities of DM into a Transformer-based framework for whole-body low-field MRI reconstruction. The proposed method exploits global feature priors to facilitate cross-contrast correlation learning and employs a DM-guided attention mechanism to recover fine-grained structural details. By generating informative 1D vector and directing attention in the frequency domain, DFAM achieves a coarse-to-fine representation that is particularly effective for reconstructing images under low-SNR conditions.

**Results:** Experimental results demonstrate that DFAM delivers high-quality reconstructions, significantly enhancing the efficiency and accuracy of low-field MRI reconstruction while ensuring the robustness of the model required for reconstructing whole-body low-field MRI.

**Conclusion:** By integrating the generative strengths of diffusion models with the representation capabilities of frequency-domain attention, DFAM effectively enhances reconstruction performance under low-SNR conditions. Experimental results demonstrate that DFAM consistently outperforms both conventional reconstruction algorithms and recent learning-based approaches. These findings highlight the potential of DFAM as a promising solution to advance low-field MRI reconstruction, particularly in resource-constrained or underdeveloped clinical settings.

# Introduction

Magnetic resonance imaging (MRI) is a non-invasive imaging technique, which can be classified into high-field ($\geq$3.0T) and low-field (<3.0T) by main magnetic field strength. A high-field MRI scanner uses stronger magnetic fields to provide a higher resolution and signal-to-noise ratio (SNR), enabling clearer views of subtle body structures and lesions. It not only has strict requirements for site conditions but also sets high standards for the professional qualities of technical personnel. Moreover, the operating cost is also high [1]. Additionally, some special effects may occur during the examination, such as radiofrequency heating, etc., for which corresponding safety measures need to be taken. Although high-field MR images are better quality, high-field equipment is difficult to popularize due to its high cost [2]. In the most developing countries, low-field MRI scanners are gradually becoming a more accessible alternative due to their high cost-effectiveness. Low-field MRI has already shown promising prospects in diagnosing brain diseases such as stroke and hydrocephalus. At the same time, the radiofrequency pulses of low-field MRI deposit less energy in tissues, which improves the safety of the scanning [3].

Despite its favorable cost-effectiveness, low-field MRI suffers from intrinsically low signal intensity due to the reduced magnetization of nuclear spins under weak magnetic fields. To acquire sufficient signal strength for clinically usable imaging, extended scan durations are often required to compensate for the inherently low signal-to-noise ratio (SNR). This reliance on prolonged acquisitions has become a major bottleneck, limiting the clinical adoption of low-field MRI and often relegating it to a supplementary or last-resort modality in diagnostic workflows [4]. To address this challenge, k-space under-sampling followed by algorithmic reconstruction has become a widely adopted strategy for accelerating MRI acquisition. Numerous studies [5–12] have demonstrated the feasibility and efficacy of this approach from various technical perspectives. However, under high acceleration factors, conventional parallel imaging techniques are prone to additional SNR degradation and residual aliasing artifacts, constraining further improvements in image quality [13]. Moreover, the compounded effects of low SNR and under-sampling artifacts hinder the reliable depiction of fine anatomical structures and pathological features, ultimately compromising diagnostic accuracy [14]. Therefore, the development of efficient and robust reconstruction methods is critical for enhancing the clinical utility of low-field MRI.

Most advancements in low-field MRI have historically relied on reconstruction and analysis techniques originally developed for high-field MRI systems. However, these conventional methods often struggle to accurately extract and restore information from low-SNR, artifact-prone data typical of low-field acquisitions, leading to suboptimal image quality and potential reconstruction errors. To address these limitations, there is an urgent need to investigate alternative image generation paradigms that can enhance both reconstruction accuracy and acquisition efficiency, thereby improving the clinical viability of low-field MRI.

Recently, diffusion model (DM) has strong representational and generative capabilities, demonstrating excellent performance in high-field MRI reconstruction tasks. For example, Tu et al. [15] trained a DM based on unsupervised scoring for MRI reconstruction on specific data. Korkmaz et al. [16] proposed SSDiffRecon, which is an MRI self-supervised DL reconstruction DM with progressive noise function. These methods demonstrate their great potential in the field of reconstruction under high-field strength. Traditional DM performs well in the image generation tasks. However, its efficiency is relatively low when applied directly to image restoration tasks [17]. In addition, compared to high-field reconstruction tasks, low-field reconstruction tasks face greater challenges, as their longer scan times and lower SNR hinder the direct application of DM in low-field reconstruction tasks [18]. Therefore, exploring the full potential of the DM in accelerating low-field MRI and improving its SNR provides a new method for low-field MRI reconstruction.

In this study, a diffusion-assisted frequency attention model (DFAM) is proposed for the reconstruction of whole-body low-field MRI. Based on the powerful generative ability of DM, we posit that the squeeze-and-excitation feature extraction (SEFE) module and the frequency attention (FA) module should be introduced to further restore fine and prominent information. The SEFE module is utilized to fuse data and extract a 1D feature vector. In lieu of traditional data concatenation, the convolutional layer was utilized to profoundly extract the global prior of low-field MR images. Subsequently, the global feature prior information undergoes compression, a process that is advantageous to the rapid convergence of DM. This approach not only ensures the high-quality generative ability of the DM but also accelerates the training of the model. The FA employs wavelet transform to convert the extraction of spatial domain features by the attention mechanism into the extraction of frequency features. Subsequently, frequency clues are utilized to explore the dependencies of latent features, thereby facilitating the comprehensive characterization of the correlations of a broader range of frequency features. The extraction of features in the frequency domain has been demonstrated to be effective in the elimination of interference from spatial domain noise, facilitating a more comprehensive capture of frequency features across different frequency domains. The main contributions are as follows:

- A SEFE module is introduced to encode global structural priors and accelerate the convergence of the DM. On one hand, the module integrates convolutional fusion of input features to extract global contextual information, effectively capturing modality-specific characteristics and cross-contrast complementarities. On the other hand, the extracted prior is compressed into a fused feature vector, enabling the diffusion process to evolve in a 1D latent space rather than traditional 2D domains. This dimensionality reduction merely requires four reverse iterations to synthesize high-quality feature representations.

- A novel FA module is introduced to effectively utilize multi-frequency features, thus mitigating noise interference and signal bias in the spatial domain. The extraction of frequency information enables more efficient encoding of image content and enhances the representation of fine-grained details. Furthermore, it achieves adaptive feature selection through a frequency-domain attention mechanism, whereby low-frequency components are used to capture overall structures, while high-frequency components focus on the refined reconstruction of textures and edges. This frequency-domain guided hierarchical learning strategy has been shown to significantly improve the model's reconstruction robustness for complex anatomical structures, particularly demonstrating superior edge-preserving capabilities under low SNR conditions.

## Theory

Low-field MRI presents unique challenges distinct from those encountered at high-field strengths [19]. Prolonged acquisition times and substantially reduced SNR limit the direct applicability of many state-of-the-art high-field reconstruction techniques in low-field scenarios [20]. Although DMs have shown remarkable capabilities in generative modeling and image denoising, their direct application to low-field MRI reconstruction remains suboptimal due to intrinsic limitations in handling complex noise patterns and signal degradation in the spatial domain. Therefore, the development of a reconstruction framework that is both noise-resilient and computationally efficient is imperative for advancing low-field MRI [21]. In this work, we propose a novel DFAM to enhance the robustness of low-field MRI reconstruction by shifting feature learning from the spatial domain to the frequency domain. The core idea is to mitigate the adverse effects of low SNR and signal heterogeneity through frequency-domain feature modulation. Specifically, DFAM incorporates two key components: (1) the SEFE module enables effective learning of the mathematical distributions of deep features, thereby capturing complex image priors with enhanced discriminative power; and (2) the FA module embeds wavelet transforms into an attention mechanism to generate frequency-aware representations that are critical for restoring fine anatomical structures under low-field conditions. Moreover, a coarse-to-fine frequency attention mechanism is designed to utilize the features among multiple frequency channels for reconstruction under the guidance of DM. To facilitate a clear understanding of the proposed framework DFAM, we begin by reviewing the forward model of low-field MRI to discuss the issue of existing methods. Building upon this theoretical foundation, we then provide a detailed description of the SEFE and FA modules, elucidating the role of each component. Finally, the training procedure and inference workflow of DFAM are systematically presented.

## A. Formulation of Low-field MRI

The mathematical model for MRI acquisition can be expressed as:

$$y = Px + n \tag{1}$$

where $y$ represents the multi-coil k-space samples collected, $x$ represents the target MR image, and $n$ is the noise. Due to the low magnetic field strength, the magnetization vector of atomic nuclei is small, resulting in weak signal intensity. The poor signal directly leads to relatively significant noise and its impact is more prominent than that in traditional high-field MRI. $P$ represents the encoding operator, which in the parallel imaging scenario is expressed as:

$$P = MFS \tag{2}$$

where $M$ is the under-sampling mask, $F$ represents the Fourier transform, and $S$ corresponds the coil sensitivity. Imaging can be accelerated and the image quality improved by collecting signals simultaneously using multiple coils and combining the sensitivity information of each coil.

However, the linear system in the formulas is underdetermined. MRI reconstruction is an ill-posed inverse problem. To obtain a high-quality reconstructed image, additional prior information is required to regularize the solution:

$$\min_x \| Px - y \|_2^2 + \alpha H(x) \tag{3}$$

where $\| Px - y \|_2^2$ denotes the data fidelity term, $H(x)$ represents the regularization term, $\| \bullet \|_2^2$ is the $l_2$ norm and $\alpha$ is the weight coefficient balancing of the regularization term. Most existing low-field MRI reconstruction approaches directly apply prior estimation strategies developed for high-field MRI, without adequately addressing the substantial SNR degradation inherent to low-field acquisitions. To mitigate this limitation, we propose an enhanced DM that extracts prior information in the frequency domain, thereby effectively suppressing spatial-domain noise interference during the prior learning process.

## B. Model Training Strategy and Variational Inference Framework

The overall DFAM framework can be divided into two sequential modes: (1) **Preliminary Training of FA Module** and (2) **Subsequent Training of Vector DM**. Central to this two-mode training strategy is the SEFE module, which acts as a critical interface between the FA and DM components. As illustrated in Fig. 1, the SEFE module initially performs the hierarchical integration of the ground-truth (GT) and the low-quality (LQ) images, leveraging a multi-scale feature extraction strategy to derive compact 1D feature vector. On one hand, the SEFE module enables the FA module to capture deep structural dependencies within MR images through encoding cross-scale contextual information. On the other hand, the SEFE module supplies pre-extracted feature vectors to DM, facilitating its learning of the underlying mathematical distribution of image features through a variational inference framework. This two-stage training strategy ensures that the model can effectively generalize to low-field MRI datasets with inherent signal noise and structural ambiguity.

**SEFE Module for Feature Fusion.** In clinical applications, the severe loss of detailed textures has always been a major issue restricting the effect of low-field MRI reconstruction. To enhance the ability of the DFAM to learn detailed textures, we design the SEFE module to obtain the fused feature vectors of images, as shown in Fig. 1(d). Specifically, this model consists of two parts. One part is the squeeze-and-excitation (SE) block, the other part is the feature extraction (FE) block. First, we input the GT and the LQ into the SE module for excitation to amplify the feature differences between the image to be reconstructed and the target image, where the GT and LQ can be denoted as $x_{GT}$ and $x_{LQ}$, respectively. In the SE block, $x_{GT}$ and $x_{LQ}$ are first converted by wavelet transform from pixel domain to frequency domain and then squeezed by the global average pool. The input data size of the $i$-th channel is $w_i \times h_i$, the global average pool can be expressed as:

$$Z_j = \frac{1}{w_{\mathbb{W},i} \times h_{\mathbb{W},i}} \sum_{m=1}^{w_{\mathbb{W},i}} \sum_{n=1}^{h_{\mathbb{W},i}} \mathbb{W}(x)_{(j,m,n)} \tag{4}$$

where $Z_j$ is the variable after the squeezing operation, $\mathbb{W}(\bullet)$ is the wavelet transform operation, $\mathbb{W}(x)_{(j,m,n)}$ represents the value at coordinates $(m,n)$ of the $j$-th channel, $w_{\mathbb{W},i}$ corresponds the width and $h_{\mathbb{W},i}$ corresponds the height of the $i$-th channel data after wavelet processing. To effectively explore the disparities between images of varying qualities and learn the interdependence among channels, the excitation operation in the network uses the convolutional layer, the ReLU and sigmoid activation functions. This process can be mathematically represented as:

$$m_j = \left( \max \left( 0, Z_j \bullet M_1 + b_1 \right) \right) \bullet M_2 + b_2 \tag{5}$$

where $m_j$ denotes the final output of the excitation operation, $M_1$ and $M_2$ are the weight matrices of the first and second convolution operations, respectively. $b_1$ and $b_2$ are the biases of the first and second convolution operations, respectively. Unlike the general sequential or simplified processing logic of typical modules, the forward-processing of SE block is as follows:

$$f = \sum_{j=1}^{C_i} \sigma(m_{GT})_j x_{GT,j} + \sum_{j=1}^{C_i} \sigma(m_{LQ})_j x_{LQ,j} \tag{6}$$

where $f$ represents the feature fusion mapping, $\sigma$ is the sigmoid function and $C_i$ is the number of channels. It aggregates the feature maps of each channel of $x_{GT}$ and $x_{LQ}$ after excitation adjustment, realizing the fusion of multi-source features and providing richer and more discriminative information for subsequent processing.

To further enhance the robustness of DFAM and enable it to better capture the diverse features of whole-body MR slices, we utilize the sliding calculation of convolutional kernels to capture local feature information such as edges and textures in the images. Meanwhile, the diversity of convolutional kernels allows for perceiving image features from multiple perspectives. Finally, the FE block integrates the feature maps to obtain the fused feature vector:

$$A = \sigma\left(\mathcal{W}_l \cdot \text{Pool}\left(\mathbb{P}(f) + \mathbb{R}(\mathbb{P}(f))\right)\right) \tag{7}$$

where $A$ denotes the extracted feature vector, $\mathbb{P}(\bullet)$ is the PixelUnshuffle operation, $\mathbb{R}(\bullet)$ is the residual transformation with 3×3 convolutions, $\mathcal{W}_l$ is the linear layer and $\text{Pool}(\bullet)$ compresses spatial dimensions.

**Preliminary Training of FA Module.** Leveraging the rich anatomical correlations among different regions in MR images, the FA module comprehensively probes the spectrum of each feature channel via frequency cues under the guidance of feature vector $A$. This grants DFAM a more potent feature-capturing capability, as depicted in Fig. 2(a). The FA module first enhances the discriminative features across each channel under the guidance of $A$, to strengthen the capability to perceive global information. It can be expressed as:

$$P' = \mathcal{W}_l A \odot \text{Norm}(P) + \mathcal{W}_l A \tag{8}$$

where $P$ and $P'$ represent the input and output feature maps, $\odot$ represents element-wise multiplication and $\text{Norm}(\bullet)$ is the normalization operation.

Following the enhancement of the discriminative features of each channel, the Transformer is employed to facilitate the recovery of fine-grained detail information. The spatio-temporal attention in the traditional Transformer can capture the dependencies between pixels. By calculating the global correlations among features, it achieves the integration of cross-regional information, which can be described as:

$$\text{Attention}(Q, K, V) = \text{softmax}\left(\frac{QK^\top}{\sqrt{d_k}}\right)V \tag{9}$$

where $Q, K, V$ correspond the features in the original space and $d_k$ is the scaling factor to prevent gradient vanishing. However, the ability of spatial-temporal attention to effectively integrate fine-grained details across channels is limited. Inspired by the structural differences of wavelet transform, we incorporate wavelet transform into the encoder-decoder architecture, enabling the attention mechanism to capture frequency feature information. The encoder-decoder architecture is shown in Figs. 2(b)-(c). Under the incorporation of wavelet transform, Eq. (9) can be redefined as:

$$WaveAttention(P') = \text{softmax}\left(\frac{(\mathcal{W}_Q \bullet \mathbb{W}(P'))(\mathcal{W}_K \bullet \mathbb{W}(P'))^\top}{\sqrt{d_k}}\right)(\mathcal{W}_V \bullet \mathbb{W}(P')) \tag{10}$$

where $\mathcal{W}_Q, \mathcal{W}_K, \mathcal{W}_V$ denote learnable linear transformation matrices. The wavelet transform and frequency attention work in tandem to form a structure-content dual-focusing mechanism. This dual-focusing architecture can accurately restore local fine structures while preserving the overall content information of the image, making it suitable for the restoration requirements of complex tissue structures in low-field MR images. The loss function is defined as follows:

$$\mathcal{L}_{TrW} = \left\| \mathbb{W}(x_{GT}) - \mathbb{W}(x_{HQ}) \right\|_1 \tag{11}$$

where $x_{HQ}$ denotes the reconstructed image and $\|\bullet\|_1$ denotes the $L_1$ norm.

**Subsequent Training of Vector DM.** The training process of DM is illustrated in Fig. 1(b), with its core lying in providing prior guidance by learning the mathematical distribution of multi-scale feature fusion vectors. Specifically, this process takes the multi-scale fused features of low-field MR images as input and models the probability distribution of feature vectors in the latent space through the reverse denoising mechanism unique to diffusion models.

During the forward training process of vector DM, we first input $x_{GT}$ and $x_{LQ}$ into the SEFE to obtain the feature vector $A$. Then, iterative noise is added to $A$ via forward DDPM, transforming their distribution into Gaussian noise $A_T \sim \mathcal{N}(0, E)$, which can be described as:

$$q(A_T \mid A) = \mathcal{N}(A_T; \sqrt{\bar{\alpha}_T} A, (1 - \bar{\alpha}_T) E) \tag{12}$$

where $\mathcal{N}$ represents the Gaussian distribution, $\alpha_T$ and $\bar{\alpha}_T$ are hyperparameters that control the variance of the noise and $E$ is the identity matrix. The loss function for training the vector DM can be expressed as follows:

$$\mathcal{L}_{diff} = \frac{1}{4C} \sum_{i=1}^{4C} |\hat{A}(i) - A(i)| \tag{13}$$

where $C$ denotes the number of channels in the SEFE module, $\hat{A}$ denotes the final reasoning result of the DM. To fully unlock the potential of DFAM in low-field MRI reconstruction, we employ a deep collaborative training strategy to optimize the DM and FA module in an end-to-end manner. This architectural design leverages a bidirectional information flow interaction mechanism to further harness the complementary strengths of generative capabilities and feature representation, thereby enhancing the reconstruction performance of DFAM. The loss function of joint training can be expressed as:

$$\mathcal{L}_{whole} = \mathcal{L}_{TrW} + \mathcal{L}_{diff} \tag{14}$$

This joint training strategy has the capacity to capture the structural correlations present at different scales. Furthermore, it can optimize the latent distribution of feature vectors through variational inference, thus rendering it more analogous to the statistical characteristics of real images.

## C. Low-field MRI Reconstruction

The reconstruction process of DFAM is illustrated in Fig. 1(c). The vector DM rapidly models the complex distribution characteristics of low-field MRI feature vectors through a stepwise denoising generation process. Guided by the feature vector generated from DM, the FA module dynamically adjusts the weights of different frequency components via an attention mechanism, emphasizing the restoration of high-frequency anatomical structures through the fusion of multi-frequency features.

In the reverse process of DM, a progressively optimized strategy is adopted based on Gaussian noise initialization. Specifically, we first generate initial noise samples following the $A_T \sim \mathcal{N}(0, E)$ distribution, then perform theoreti-cally guaranteed iterative denoising through the DM. It leverages the information $\eta_\theta(A_t, t, U)$ in the forward-process and is assisted by the feature vector $U$ extracted from the LQ low-field MRI data. This iterative sampling continues until $t = 1$, at which point the final generated result $\tilde{A}$ is obtained. This sampling process can be described as follows:

$$\tilde{A}_{t-1} = \frac{1}{\sqrt{\alpha_t}}(A_t - \frac{1-\alpha_t}{\sqrt{1-\bar{\alpha}_t}}\eta_\theta(A_t, t, U)) \tag{15}$$

After obtaining the feature fusion vector $\tilde{A}$, the pre-trained FA module can perform high-quality reconstruction of the under-sampled images. This process leverages the abundant contextual information embedded in $\tilde{A}$ to dynamically adjust the attention weights across different frequency components, ensuring that both the global structural consistency and local fine details of the images are preserved.

During MRI reconstruction, the occurrence of data mismatches between reconstructed images and acquired k-space measurements is a frequent phenomenon. To guarantee enhanced image reliability, data consistency (DC) operations are implemented with a view to eliminating potential artefacts arising from data mismatches. DC operation can be described as:

$$K_j = \begin{cases} (K_{HQ})_j, & \text{if } j \notin \Omega \\ \dfrac{(K_{HQ})_j + \lambda(K_{LQ})_j}{(1+\lambda)}, & \text{if } j \in \Omega \end{cases} \tag{16}$$

where $\Omega$ is the index set of the collected k-space samples, $K_{LQ}$ corresponds the collected k-space samples, $K_{HQ}$ corresponds the value of the image k-space data generated by the model at index $j$, and $\lambda$ is the weight coefficient, which is used to balance the impact of spatial measurement noise on reconstruction. If the $j$-th prediction data has been sampled without noise (i.e. $\lambda \to \infty$), it will be replaced by the original data. The overall reconstruction process of DFAM module is presented in algorithm 1 of **Appendix A**.

## Method

**Description of Datasets.** We use the brain images from SIAT and TotalSegmentator whole-body images [22] as the training dataset. The SIAT dataset from the Shenzhen Institutes of Advanced Technology has obtained ethical approval and contains 500 12-channel MR images of healthy volunteers. Acquired by 3.0 T Siemens Trio Tim with T2-weighted turbo spin echo (repetition time (TR)/echo time (TE)=6100/99 $ms$, $220 \times 220$ $mm^2$ field of view), they are merged and augmented to 4000 single-channel images. TotalSegmentator offers unprecedented data scale and diversity to enable better adaptation to variable clinical scenarios. The dataset contains 298 MR images. Among them, 251 MR images are from the Picture Archiving and Communication System of the University Hospital Basel between 2011 and 2023, and the other 47 MR images are from the Imaging Data Commons. High-quality images are under-sampled (3×-12×) to simulate low-field MRI data. The high-quality original images and their corresponding low-quality simulated under-sampled images are combined into paired data for training.

To verify the generalization performance of the DFAM, we test single-channel MR images that had undergone coil fusion and are acquired by a 0.3T low-field MRI device from Shenzhen Anke High-Tech Co., Ltd. There are a total of five image datasets scanned from different parts of the human body, namely head, cervical spine, abdomen, lumbar spine, and knee joint. By testing the low-field MR images of different body parts, we tend to demonstrate the effectiveness and generalization ability of the proposed DFAM.

**Parameter Configuration.** The DFAM is trained using the Adam optimizer with momentum terms set to $\beta_1$=0.9 and $\beta_2$=0.999. Additionally, we employ a learning rate scheduler to dynamically adjust the learning rate, leveraging cyclic restarts to allow the model multiple opportunities to explore different learning rate configurations. Within each cycle, the learning rate gradually decreases, enabling the model to perform fine-tuning in later stages. In the FA module, the wavelet transform method employed is the Daubechies 1 Extremal Phase Wavelet Transform. In the Transformer modules, the number of Transformer blocks is set to be [3, 5, 6, 6]. The number of attention heads is set to be [1, 2, 4, 8]. The input and output channel dimensions are 4. For the DM, the number of iterations is set to 4. To better explore the reconstruction ability of the DFAM, both the training size and the batch size vary with the increase of training epochs. All experiments were conducted in the PyTorch environment with an NVIDIA RTX4090D GPU. The source code is available at: https://github.com/yqx7150/DFAM.

**Evaluation Metrics.** To quantitatively assess the fidelity of the reconstructed images relative to the reference targets, we employ three standard image quality metrics: mean square error (MSE), peak signal-to-noise ratio (PSNR), and structural similarity index measure (SSIM). In order to further evaluate the reconstruction efficiency of the proposed DFAM framework in the context of low-field MRI, we introduce a novel composite metric termed the Efficiency-Quality Ratio (EQRatio). This metric jointly considers image quality—captured by PSNR and SSIM—and reconstruction latency. The quality component is defined by the improvement in PSNR and SSIM from the under-sampled input to the final reconstructed image. Notably, the influence of reconstruction time on perceived image quality is inherently non-linear [23]; as reconstruction time increases, its marginal contribution to perceived quality improvement diminishes. To account for this non-linearity, a logarithmic transformation is applied to the reconstruction time prior to integration into the EQRatio metric formulation, thereby enabling a more realistic and robust assessment of reconstruction efficiency. The higher the EQRatio value, the higher the reconstruction efficiency. The EQRatio is defined by:

$$EQRatio = \frac{\omega_1 \times (PSNR_{Rec} - PSNR_{Under}) + \omega_2 \times (SSIM_{Rec} - SSIM_{Under})}{\text{Log}(T)} \tag{17}$$

where $\omega_1$=0.1 and $\omega_2$=0.9 are the index weights, $PSNR_{Rec}$ and $PSNR_{Under}$ corresponds the PSNR of the reconstructed state and the PSNR in the under-sampled state, $SSIM_{Rec}$ and $SSIM_{Under}$ corresponds the SSIM of the reconstructed state and the SSIM in the under-sampled state, $T$ is the time required for reconstruction, with the unit being second.

# Results

## A. Comparison with State-of-the-arts

To validate the effectiveness of the proposed DFAM framework, its performance is quantitatively and qualitatively evaluated against several state-of-the-art reconstruction methods. These include traditional model-based approaches such as P-LORAKS [24] and ESPIRiT [25], as well as deep learning–based methods including EBMRec [26] and MoDL [27]. Comparative experiments are conducted under a variety of sampling strategies—Poisson sampling, ran-dom sampling, and uniform sampling—across multiple acceleration factors (i.e., 3×, 4×, 5×, 6×, and 10×), thereby enabling a comprehensive assessment of robustness and generalizability under diverse under-sampling conditions.

Figs. 3 and 4 illustrate the reconstruction results of low-field MRI of the lumbar and cervical vertebrae. As expected, the frequency attention mechanism can more accurately distinguish and preserve the detailed textures of low-field MR images. The error maps show that traditional algorithms perform poorly, which strongly verifies the negative impact of the weak signal characteristics of low-field MRI on the reconstruction results of traditional algorithms. This is because the T1/T2 relaxation times change with the field strength, and traditional high-field reconstruction algorithms cannot effectively distinguish the real signal from noise under low SNR, leading to varying degrees of degradation in reconstruction performance. As can be seen from the error maps in Fig. 4, when the DL algorithms are directly transferred to low-field conditions, due to the differences in signal statistical distribution, residual artifacts still exist in the reconstruction results of EBMRec and MoDL. Table 1 presents the quantitative reconstruction results of the lumbar and cervical vertebra datasets, with the best results indicated in bold. The quantitative results demonstrate the effectiveness of the proposed method, showing that it consistently outperforms other methods across different sampling patterns.

## B. Comparison with DM-based Methods

To validate the effectiveness of transforming the diffusion modeling target from a 2D image to a 1D vector representation, we compare the proposed DFAM framework with two diffusion model–based MRI reconstruction methods Score-MRI [28] and HGGDP [29]. In both baseline methods, the forward diffusion process involves the progressive corruption of a fully sampled 2D MR image by Gaussian noise until it becomes nearly pure noise. During the reverse process, a trained generative model is employed to iteratively remove the noise, thereby reconstructing the high-quality MR image from the degraded input.

Fig. 5 presents the comparison results of the reconstruction quality of head low-field MRI among DFAM, Score-MRI, and HGGDP. The experimental evaluations are conducted under the 1D uniform sampling mode with acceleration factors $R$=4, 8, and 2D random sampling mode with acceleration factors $R$=8, 12, respectively. The error maps show that DFAM performs better in noise removal, and the reconstruction results eliminate more artifacts. Moreover, in the marked region of interest, DFAM demonstrates superior preservation of detailed textures, enabling clearer visualization of brain scan details. Table 2 tabulates the quantitative comparison results, where the best results are marked in bold black font. This demonstrates that DFAM not only performs excellently at low acceleration factors but also achieves outstanding results as the acceleration factor increases.

## C. Comparison with Transformer-based Methods

The goal of DFAM is to implicitly guide the attention model to focus on learning structural-relevant features in the frequency domain. To this end, we first exam the effectiveness of DFAM by comparing it with SwinIR [30] and Restormer Transformer [31]. It should be emphasized that both SwinIR and Restormer Transformer directly process the input degraded low-field MR images in the image domain. They capture features in the image domain through channel attention, and their potential limitation is that the models may be more inclined to learn the structural information of images while neglecting the preservation of detailed textures.

Fig. 6 depicts the comparison results of DFAM with SwinIR and Restormer. This experiment is conducted on the low-field MRI knee slice dataset, using 1D uniform sampling mode with acceleration factors $R$=4, 5, and 2D Poisson sampling mode with acceleration factors $R$=4, 6. As can be seen from the error maps, DFAM demonstrates unique advantages through its frequency-domain learning mechanism. The low-frequency channel ensures the accuracy of anatomical structures, while the high-frequency channel focuses on texture detail enhancement. This enables DFAM to significantly excel in artifact removal across the entire image and achieve more distinct reconstruction of image boundary information. Table 3 shows the comparison of quantitative indicators between DFAM and the two comparison methods, and the best reconstruction results are marked in bold black. Our method achieves better reconstruction effects under different knee slices and different sampling multiples, which further illustrates the effectiveness and robustness of low-field MRI reconstruction in the frequency domain.

## Discussion

Through the above experiments, we have demonstrated the effectiveness of DFAM for the low-field MRI reconstruction in the frequency domain. Next, we will explore the optimization capabilities of different innovative modules for DFAM and compare the efficiency of our method with that of other reconstruction methods, to analyze its overall advantages and potentials in the low-field MRI reconstruction.

### A. Ablation Study on the Model Architecture

Our work builds upon the foundation of DiffIR [17]. However, direct application of the original DiffIR model to the reconstruction of low-field MR images yields suboptimal results. To address this limitation, we propose a series of targeted modifications and enhancements to the DiffIR architecture. Specifically, we reformulate the reconstruction pipeline by transitioning the processing domain from the image domain to the frequency domain. In addition, an image feature fusion mechanism is introduced to facilitate the extraction of more salient structural and textural features. To rigorously assess the efficacy of the proposed improvements, we perform extensive comparative experiments against the baseline model and other representative methods.

Supporting Information Figure S1 displays a comparison of the quantitative indicators of the reconstruction results with and without the application of the SEFE and FA modules. The best results in each experiment are marked in bold black font. The bar chart shows that the attention model with the SEFE module or FA module individually outperforms the DiffIR. Moreover, the proposed DFAM has achieved the best results. Experimental results show that the frequency-domain-based DFAM significantly improves the reconstruction effect of low-field MRI compared with traditional image domain calculations. Meanwhile, the deep fusion of input features can effectively improve the reconstruction robustness, thus better supporting whole-body low-field MRI reconstruction.

### B. Ablation Analysis of Computational Efficiency

The superior reconstruction efficiency of the proposed DFAM framework in low-field MRI arises from the synergistic design of the FA module and the integration of a vector DM. Specifically, the FA module employs an end-to-end training and reconstruction strategy that eliminates the need for iterative optimization during inference, thereby substantially reducing computational overhead. Meanwhile, the vector DM is leveraged to generate compact and informative feature vectors that capture the essential structural characteristics of the target images, enabling accurate and rapid reconstruction in an extremely short time. To quantitatively evaluate the reconstruction efficiency of DFAM, we compare its performance with three representative baseline methods: the traditional model-based algorithm SAKE [32], the end-to-end deep learning model Restormer [31], and the generative reconstruction approach WKGM [15]. Furthermore, we introduce the EQRatio as a unified metric that simultaneously accounts for image quality and reconstruction latency, where higher EQRatio values indicate superior overall reconstruction efficiency.

Supporting Information Table S1 exhibits the quantitative results of the reconstruction efficiency between DFAM and SAKE, Restormer, and WKGM. This experiment is carried out under radial sampling with $R$=8 and $R$=4. The best results are presented in bold black. From the results, DFAM can achieve better reconstruction quality and obtain the highest reconstruction efficiency simultaneously. Specifically, DFAM has an advantage in terms of faster calculation speed compared with SAKE and WKGM. Although our method has an SSIM value 0.007 lower than that of Restormer when $R$=8, reconstruction speed of DFAM is 1 second faster than that of Restormer, and the comprehensive efficiency is better. Therefore, it can be concluded that DFAM has a very high-efficiency in reconstructing low-field MR images. This characteristic enables our method to provide a better option for the reconstruction of low-field MRI in underdeveloped regions.

## Conclusion

In this study, a novel mathematical framework is proposed for whole-body low-field MRI reconstruction, which jointly exploits feature correlations and frequency-domain cues. Specifically, by integrating the powerful generative capability of DM and the multi-scale representation capability of frequency attention, we addressed the trade-off between reconstruction accuracy and computational efficiency. Furthermore, fine-grained and salient anatomical details were enhanced through the deep exploitation of global feature priors. To facilitate comprehensive performance evaluation, we also introduced a composite metric that simultaneously accounts for reconstruction quality and runtime. Both theoretical analysis and experimental results demonstrated that the proposed DFAM framework consistently outperforms existing low-field MRI reconstruction methods, offering a promising solution for resource-constrained and underdeveloped clinical settings.

## Acknowledgment

This work was supported in part by the National Key Research and Development Program of China under Grant 2023YFF1204300 and Grant 2023YFF1204302, and Key Research and Development Program of Jiangxi Province under 20212BBE53001.

## Appendix A: Diffusion-Assisted Frequency Attention Model for MRI Reconstruction

---

**Algorithm 1:** DFAM **for MRI Reconstruction**

---

**Require:** Trained fusion feature extraction module SEFE, low-field MRI reconstruction module FA, trained DM $\eta_\theta(A_t, t, \tilde{A})$, under-sampled low-field MR image $x$

**1: Initialization:** $A_T \sim \mathcal{N}(0, E)$

**2: Reconstruction image feature generation:**

    **Extracting** under-sampling image features $\tilde{A}$ by Eq. (7)

    **For** $t = 4$ **to** $1$ **do**

        **Update** $A_{t-1}$ via Eq. (14)

    **End for**

**3: Preliminary reconstruction of under-sampled image:**

    **Transform** image $x_{LQ}$ into frequency domain by wavelet transform

    **Update** $x$ via Eqs. (8), (10)

    **For** $t = 3$ **to** $1$ **do**

        **Encode** $x_t$

        **Update** $x_t$ via Eqs. (8), (10)

    **End for**

    **For** $t = 3$ **to** $1$ **do**

        **Decode** $x_t$

        **Update** $x_t$ via Eqs. (8), (10)

    **End for**

    **Convert** the result $x_{HQ}$ back to the image domain by inverse wavelet transform.

**4: Update** $x_{HQ}$ by data consistency with Eq. (16)

**5: Return** $x_{Rec} = SOS(x_{HQ})$

---

## Captions

**Fig. 1.** The training and reconstruction process of the DFAM. (a) Training of the FA module. (b) Training of the one-dimensional DM. (c) Reconstruction of the DFAM. (d) The structure of the SEFE module. This module is used to fuse the paired training data and extract feature vector from it, thereby guiding the training of the model.

**Fig. 2.** The structure of the FA module. The encoder and decoder modules in the model are designed using wavelet transform and inverse wavelet transform, respectively. After each encoding or decoding operation, the calculations are performed in the Transformer module. Through this model, the input data is reconstructed into high-quality images with the assistance of the corresponding feature vector.

**Fig. 3.** Reconstruction results of lumbar slices under $R$=6 Poisson sampling. From left to right: reference image, zero-filled result and the images reconstructed by different methods. The second row shows the enlarged view of the ROI, which is the area marked by the yellow box in the first row. The third row shows the error maps of the reconstruction results.

**Fig. 4.** Reconstruction results of cervical slices under $R$=4 uniform sampling. From left to right: reference image, zero-filled result and the images reconstructed by different methods. The second row shows the enlarged view of the ROI, which is the area marked by the yellow box in the first row. The third row shows the error maps of the reconstruction results.

**Fig. 5.** Reconstruction results of low-field head slice images under $R$=8 random sampling. From left to right: reference image, zero-filled result and the images reconstructed by different methods. The second row shows the magnified view of the reconstructed ROI. The third row shows the error maps of the reconstruction results.

**Fig. 6.** Reconstruction result images of low-field MRI knee slices under $R$=5 1D uniform sampling. From left to right: reference image, zero-filled result and the images reconstructed by different methods. The second row shows the magnified view of the reconstructed ROI. The third row shows the error maps of the reconstruction results.

**Table 1.** The PSNR, SSIM and MSE of reconstruction results under various sampling patterns and acceleration factors are compared. Test 1 uses a 0.3 T low-field lumbar spine slice and test 2 uses a 0.3 T cervical spine slice.

**Table 2.** The PSNR, SSIM and MSE of reconstruction results under various sampling patterns and acceleration factors are compared. Test 1 and test 2 involve different 0.3T low-field MRI head slices.

**Table 3.** The PSNR, SSIM and MSE of reconstruction results under various sampling patterns and acceleration factors are compared. Test 1 involves a 0.3 T low-field knee joint slice and test 2 involves a 0.3 T abdominal slice.

## List of supporting information

**Supporting Information Figure S1.** This shows the PSNR, SSIM and MSE metrics of different MRI reconstruction results under various acceleration factors. The test data comprises different 0.3 T low-field MRI abdominal slices. The methods compared are: DiffIR, SEFE, FA and DFAM.

**Supporting Information Table S1.** PSNR, SSIM, TIME and EQ ratio of reconstruction results under radial sampling patterns at various acceleration factors. The test data is a 0.3 T low-field MRI head slice.

# Figures

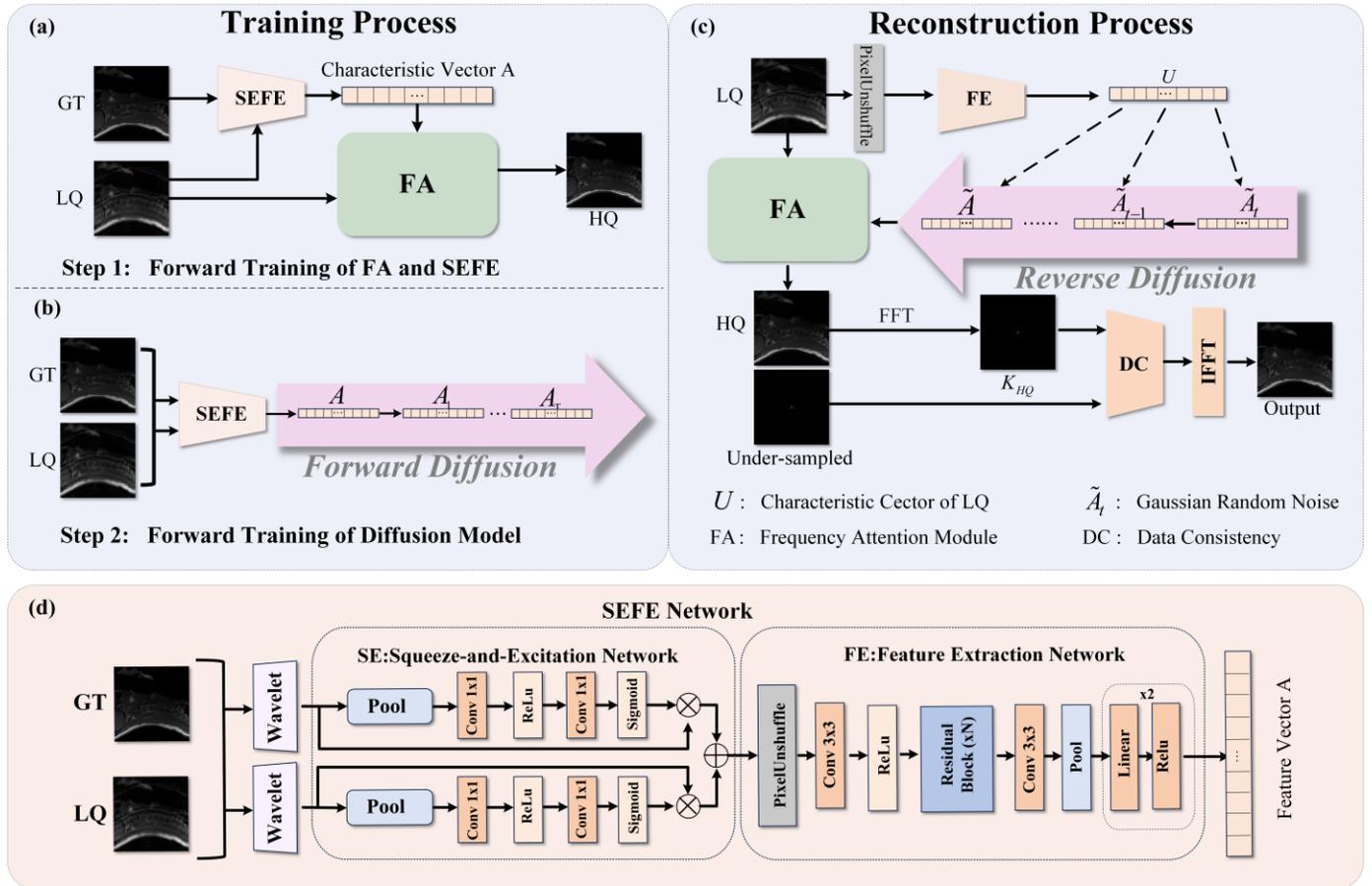

**Fig. 1.** The training and reconstruction process of the DFAM. (a) Training of the FA module. (b) Training of the 1D-dimensional DM. (c) Reconstruction of the DFAM. (d) The structure of the SEFE module. This module is used to fuse the paired training data and extract feature vector from it, thereby guiding the training of the model.

**Fig. 2.** The structure of the FA module. The encoder and decoder modules in the model are designed using wavelet transform and inverse wavelet transform, respectively. After each encoding or decoding operation, the calculations are performed in the Transformer module. Through this model, the input data is reconstructed into high-quality images with the assistance of the corresponding feature vector.

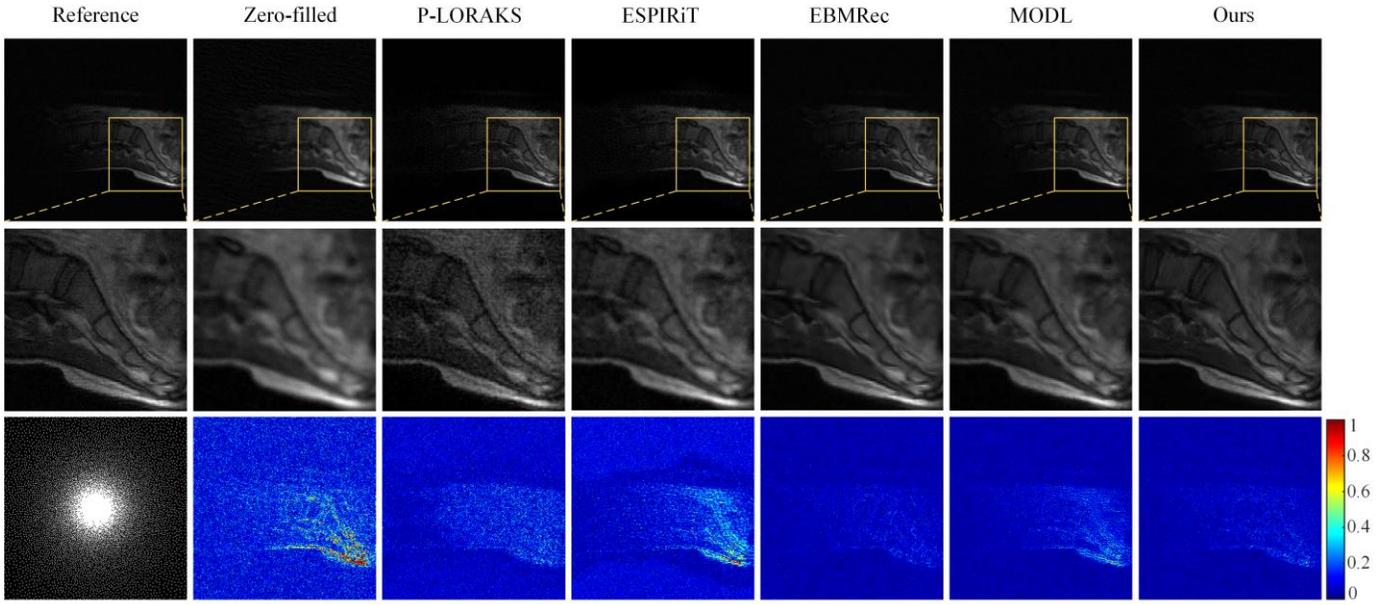

**Fig. 3.** Reconstruction results of lumbar slices under $R$=6 Poisson sampling. From left to right: reference image, zero-filled result and the images reconstructed by different methods. The second row shows the enlarged view of the ROI, which is the area marked by the yellow box in the first row. The third row shows the error maps of the reconstruction results.

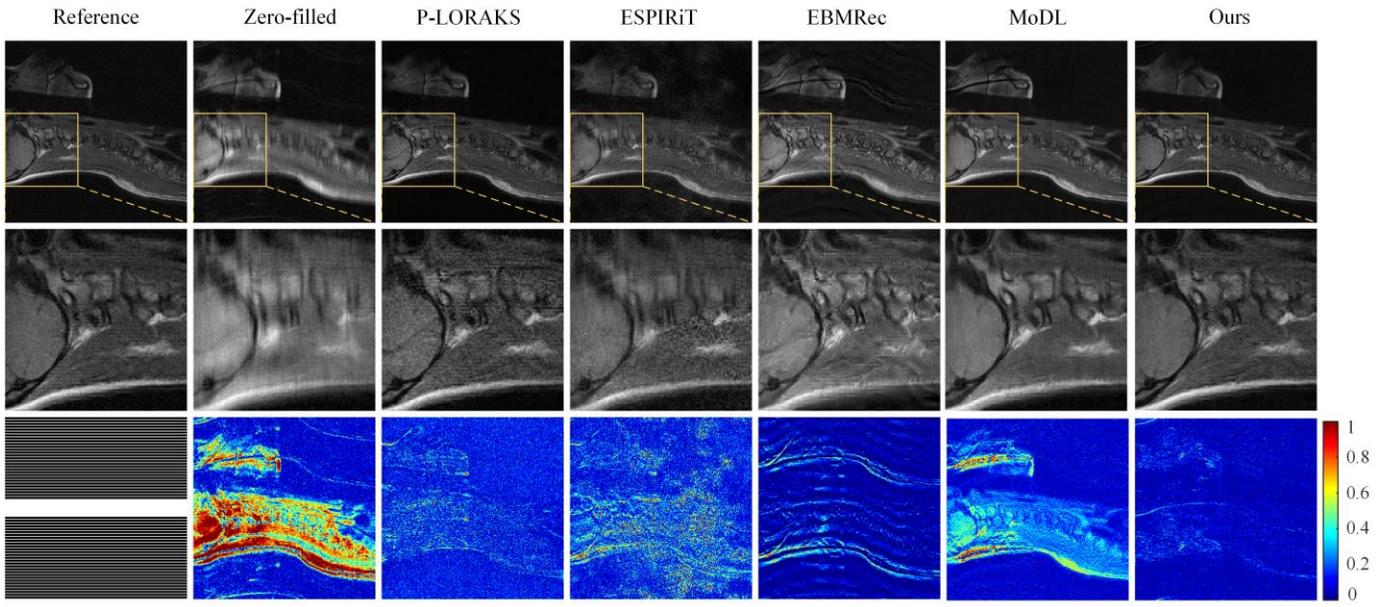

**Fig. 4.** Reconstruction results of cervical slices under $R$=4 uniform sampling. From left to right: reference image, zero-filled result and the images reconstructed by different methods. The second row shows the enlarged view of the ROI, which is the area marked by the yellow box in the first row. The third row shows the error maps of the reconstruction results.

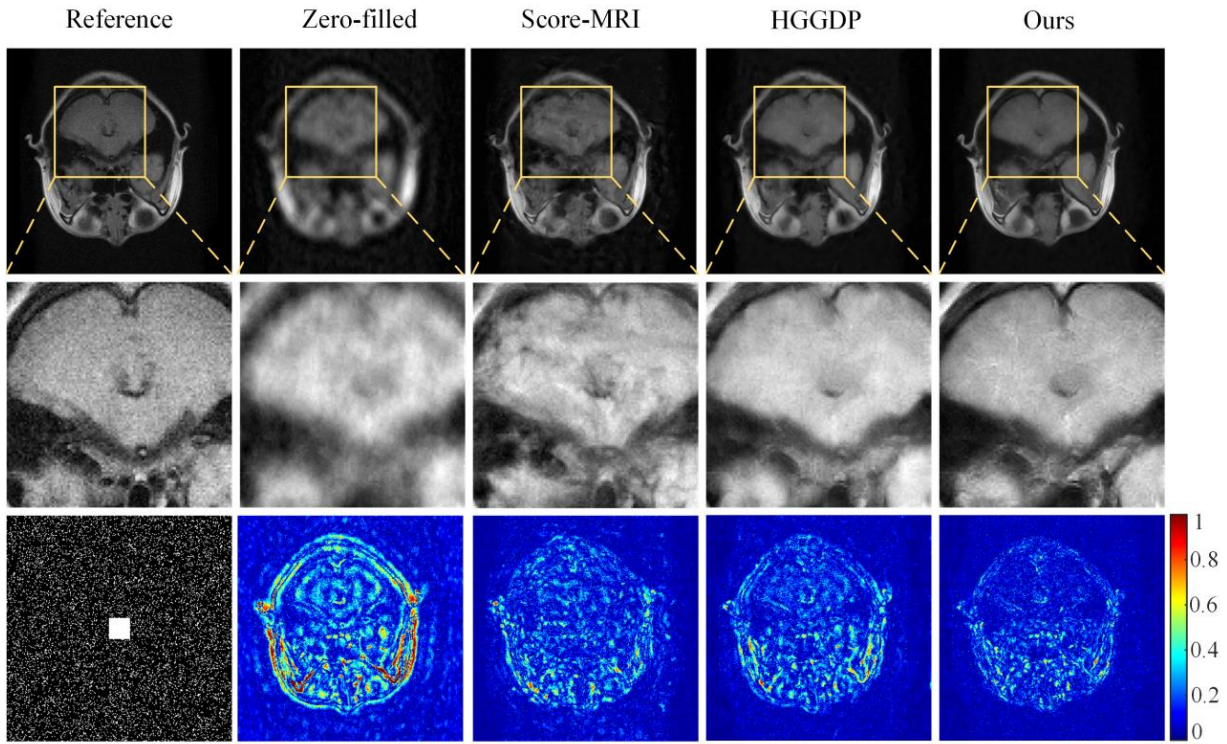

**Fig. 5.** Reconstruction results of low-field head slice images under *R*=8 random sampling. From left to right: reference image, zero-filled result and the images reconstructed by different methods. The second row shows the magnified view of the reconstructed ROI. The third row shows the error maps of the reconstruction results.

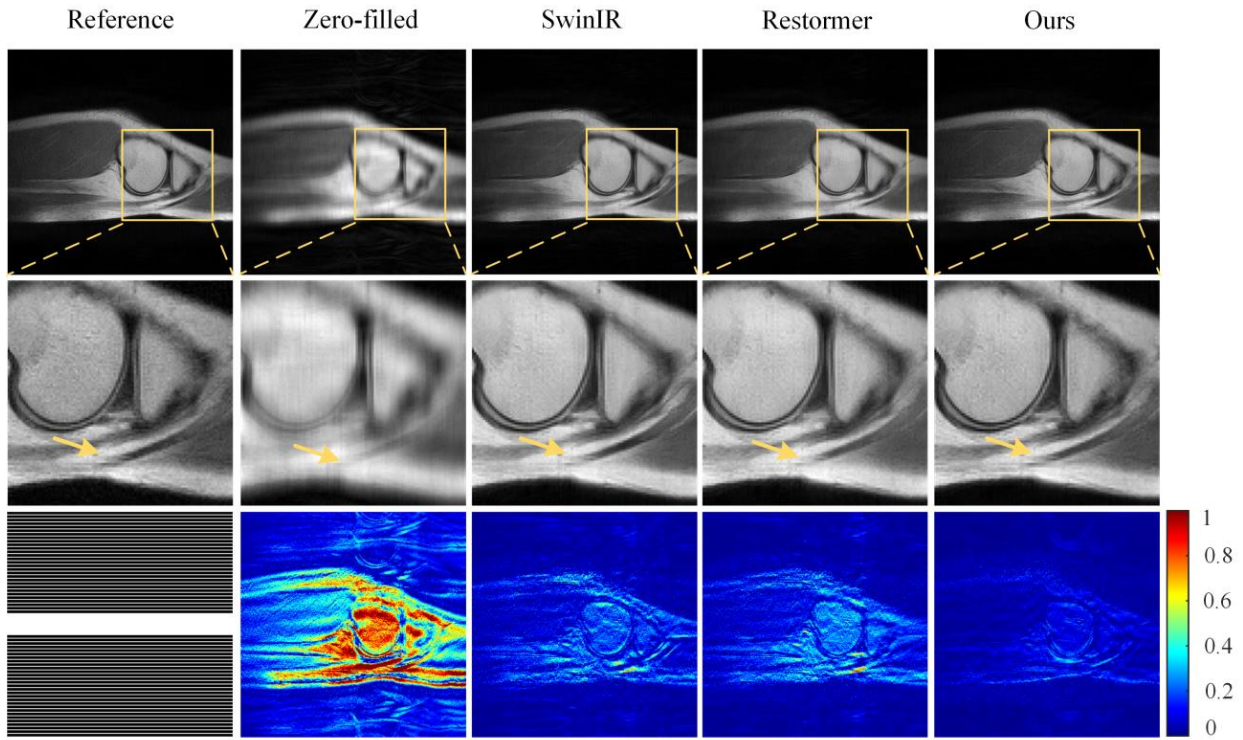

**Fig. 6.** Reconstruction result images of low-field MRI knee slices under $R$=5 1D uniform sampling. From left to right: reference image, zero-filled result and the images reconstructed by different methods. The second row shows the magnified view of the reconstructed ROI. The third row shows the error maps of the reconstruction results.

| *Test 1* | Zero-filled | P-LORAKS | ESPIRiT | EBMRec | MoDL | Ours |
|---|---|---|---|---|---|---|
| Poisson $R$=6 | 28.91/0.7957/12.84 | 33.41/0.6367/4.550 | 32.05/0.5580/6.226 | 38.21/0.8974/1.507 | 37.84/0.9065/1.643 | **38.37/0.9084/1.455** |
| Poisson $R$=10 | 24.95/0.6899/31.92 | 32.25/0.5937/5.956 | 29.40/0.5115/11.47 | 37.01/0.8780/1.986 | 36.80/0.8899/2.085 | **37.09/0.8903/1.951** |
| Random $R$=4 | 28.12/0.7819/15.40 | 31.49/0.5725/7.092 | 32.58/0.5314/5.511 | 37.89/0.9031/1.624 | 38.08/0.9094/1.552 | **38.49/0.9166/1.414** |
| Random $R$=5 | 27.43/0.7682/18.04 | 32.03/0.5833/6.258 | 31.64/0.5304/6.839 | 38.24/0.8987/1.498 | 36.88/0.8883/2.049 | **38.36/0.9098/1.456** |

| *Test 2* | Zero-filled | P-LORAKS | ESPIRiT | EBMRec | MoDL | Ours |
|---|---|---|---|---|---|---|
| Uniform $R$=4 | 19.87/0.6547/103.0 | 29.24/0.727/11.91/ | 25.77/0.5630/26.48 | 29.44/0.7930/11.37/ | 32.62/0.8714/5.460 | **34.85/0.8833/3.270** |
| Uniform $R$=6 | 18.40/0.5160/144.3 | 26.50/0.6433/22.37 | 24.27/0.4333/37.36 | 26.93/0.6602/20.24 | 27.98/0.7351/15.90 | **28.73/0.7498/13.38** |
| Random $R$=5 | 25.47/0.7029/28.32 | 28.97/0.7026/12.65 | 28.66/0.6659/13.59 | 31.75/0.7875/6.673 | 31.79/0.7936/6.621 | **31.93/0.8284/6.402** |
| Random R=3 | 27.16/0.7634/19.19 | 30.70/0.7709/8.503/ | 30.29/0.7283/9.352 | 34.44/0.8624/3.589 | 34.72/0.8755/3.371 | **34.99/0.8765/3.168** |

**Table 1.** The PSNR, SSIM and MSE of reconstruction results under various sampling patterns and acceleration factors are compared. Test 1 uses a 0.3 T low-field lumbar spine slice and test 2 uses a 0.3 T cervical spine slice.

| *Test 1* | Zero-filled | Score-MRI | HGGDP | Ours |
|----------|-------------|-----------|-------|------|
| Uniform $R$=4 | 29.59/0.8146/10.97 | 34.37/ 0.8468/3.659 | 34.29/0.8967/3.719 | **35.89/0.9196/2.575** |
| Uniform $R$=8 | 28.50/0.7770/14.12 | 31.58/0.7980/6.943 | 31.51/0.8475/7.067 | **32.35/0.8655/5.823** |
| *Test 2* | Zero-filled | Score-MRI | HGGDP | Ours |
| Random $R$=8 | 24.33/0.5528/36.89 | 28.31/0.7058/14.74 | 29.76/0.7624/10.55 | **30.33/0.7699/9.271** |
| Random $R$=12 | 23.88/0.5481/40.97 | 27.00/0.6702/ 19.93 | 27.62/ 0.6945/17.29 | **28.94/0.7587/12.76** |

**Table 2.** The PSNR, SSIM and MSE of reconstruction results under various sampling patterns and acceleration factors are compared. Test 1 and test 2 involve different 0.3T low-field MRI head slices.

| *Test 1* | Zero-filled | SwinIR | Restormer | Ours |
|---|---|---|---|---|
| Uniform *R*=4 | 21.26/0.5514/74.73 | 35.03/0.8399/3.139 | 32.25/0.8459/5.950 | **36.10/0.8998/2.450** |
| Uniform *R*=5 | 20.69/0.5199/85.22 | 30.91/0.8019/8.096 | 30.04/0.8194/9.906 | **31.71/0.8637/6.742** |
| *Test 2* | Zero-filled | SwinIR | Restormer | Ours |
| Poisson R=4 | 27.74/0.6449/16.81 | 34.60/0.8670/3.459 | 36.10/0.8807/2.454 | **36.69/0.8887/2.138** |
| Poisson R=6 | 24.54/0.5596/35.08 | 31.36/0.8317/7.308 | 33.22/0.8530/4.758 | **34.86/0.8675/3.258** |

**Table 3.** The PSNR, SSIM and MSE of reconstruction results under various sampling patterns and acceleration factors are compared. Test 1 involves a 0.3 T low-field knee joint slice and test 2 involves a 0.3 T abdominal slice.

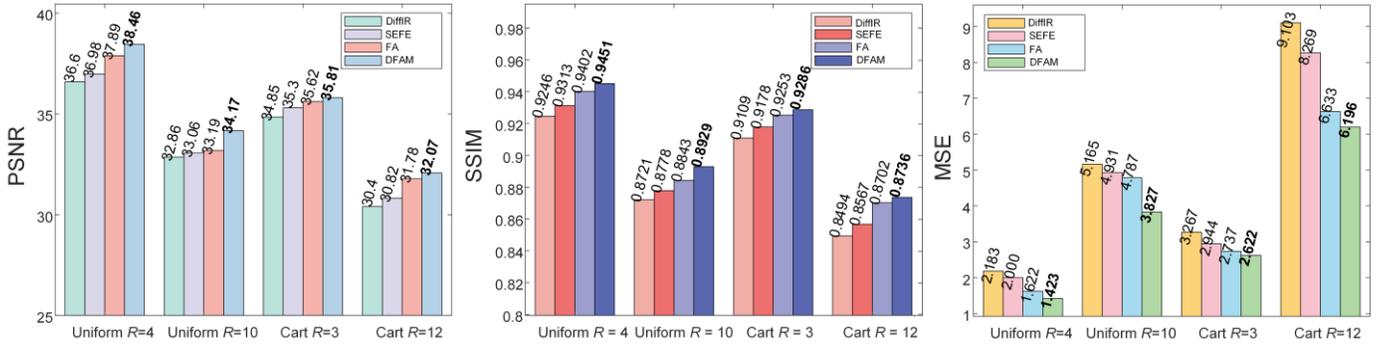

**Supporting Information Figure S1.** This shows the PSNR, SSIM and MSE metrics of different MRI reconstruction results under various acceleration factors. The test data comprises different 0.3 T low-field MRI abdominal slices. The methods compared are: DiffIR, SEFE, FA and DFAM.

| *Radial R=8* | Zero-filled | SAKE | Restormer | WKGM | Ours |
|:---:|:---:|:---:|:---:|:---:|:---:|
| PSNR | 17.78 | 29.18 | 29.46 | 29.26 | **30.26** |
| SSIM | 0.4422 | 0.6499 | **0.7794** | 0.6495 | 0.7724 |
| TIME | / | 45.06 | 4.451 | 54.45 | **3.462** |
| EQRatio* | / | 0.3485 | 0.9855 | 0.3339 | **1.2443** |
| *Radial R=4* | Zero-filled | SAKE | Restormer | WKGM | Ours |
| PSNR | 25.78 | 30.29 | 35.08 | 32.10 | **35.81** |
| SSIM | 0.6516 | 0.6834 | 0.8647 | 0.7487 | **0.8794** |
| TIME | / | 37.27 | 4.267 | 19.92 | **3.634** |
| EQRatio* | / | 0.1326 | 0.7732 | 0.2405 | **0.9362** |

**EQRatio***: The EQRatio is the efficiency-quality ratio. This indicator comprehensively considers the PSNR, the SSIM, and the reconstruction time. The higher the EQRatio value, the higher the reconstruction efficiency.

**Supporting Information Table S1.** PSNR, SSIM, TIME and EQ ratio of reconstruction results under radial sampling patterns at various acceleration factors. The test data is a 0.3 T low-field MRI head slice.